\documentclass[twocolumn]{svjour3}
\usepackage{graphics}
\usepackage{hyperref}
\usepackage{amssymb}
\usepackage{braket}
\usepackage[numbers,sort&compress]{natbib}

\usepackage{color}
\usepackage[dvipsnames]{xcolor}

\journalname{Granular Matter}

\begin{document}

\title{Enlightening force chains: a review of photoelasticimetry in granular matter}

\author{
Aghil Abed Zadeh \and 
Jonathan Bar\'es \and
Theodore A. Brzinski \and 
Karen E. Daniels \and
Joshua Dijksman \and
Nicolas Docquier \and
Henry Everitt \and
Jonathan E. Kollmer \and
Olivier Lantsoght \and
Dong Wang \and
Marcel Workamp \and
Yiqiu Zhao \and
Hu Zheng 
}                    

\institute{
Aghil Abed Zadeh, Department of Physics \& Center for Nonlinear and Complex Systems, Duke University, Durham, NC, USA \and 
Jonathan Bar\'es, Laboratoire de M\'{e}canique et G\'{e}nie Civil, Universit\'{e} de Montpellier, CNRS, Montpellier, France, \email{jb@jonathan-bares.eu} \and
Theodore A. Brzinski, Department of Physics, Haverford College, Haverford, PA, USA \and 
Karen E. Daniels, Department of Physics, North Carolina State University, Raleigh, NC, USA \and
Joshua Dijksman, Physical Chemistry and Soft Matter, Wageningen University \& Research, Wageningen, The Netherlands \and
Nicolas Docquier, Institute of Mechanics, Material and Civil engineering, Universit\'{e} catholique de Louvain, Louvain-la-Neuve, Belgium \and
Henry Everitt, Department of Physics and Department of Chemistry, Duke University, Durham, NC, USA \and
Jonathan E. Kollmer, Department of Physics, Universit\"at Duisburg-Essen, Duisburg, Germany \and
Olivier Lantsoght, Institute of Mechanics, Material and Civil engineering, Universit\'{e} catholique de Louvain, Louvain-la-Neuve, Belgium \and
Dong Wang, Department of Physics \& Center for Nonlinear and Complex Systems, Duke University, Durham, NC, USA \and
Marcel Workamp, Physical Chemistry and Soft Matter, Wageningen University \& Research, Wageningen, The Netherlands \and
Yiqiu Zhao, Department of Physics \& Center for Nonlinear and Complex Systems, Duke University, Durham, NC, USA \and
Hu Zheng, Department of Physics \& Center for Nonlinear and Complex Systems, Duke University, Durham, NC, USA; School of Earth Science and Engineering, Hohai University, Nanjing, Jiangsu, 211100, China 
}

\date{Received: date / Revised version: date}

\maketitle

\begin{abstract}
A photoelastic material will reveal its internal stresses when observed through polarizing filters. This eye-catching property has enlightened our understanding of granular materials for over half a century, whether in the service of art, education, or scientific research. In this review article in honor of Robert Behringer, we highlight both his pioneering use of the method in physics research, and its reach into the public sphere through museum exhibits and outreach programs. We aim to provide clear protocols for artists, exhibit-designers, educators, and scientists to use in their own endeavors. It is our hope that this will build awareness about the ubiquitous presence of granular matter in our lives, enlighten its puzzling behavior, and promote conversations about its importance in environmental and industrial contexts. To aid in this endeavor, this paper also serves as a front door to a detailed wiki containing open, community-curated guidance on putting these methods into practice \cite{all2019_wiki}.
\end{abstract}

\section{Introduction} \label{sec_intro} 

\begin{figure}
\centering 
\resizebox{\hsize}{!}{\includegraphics{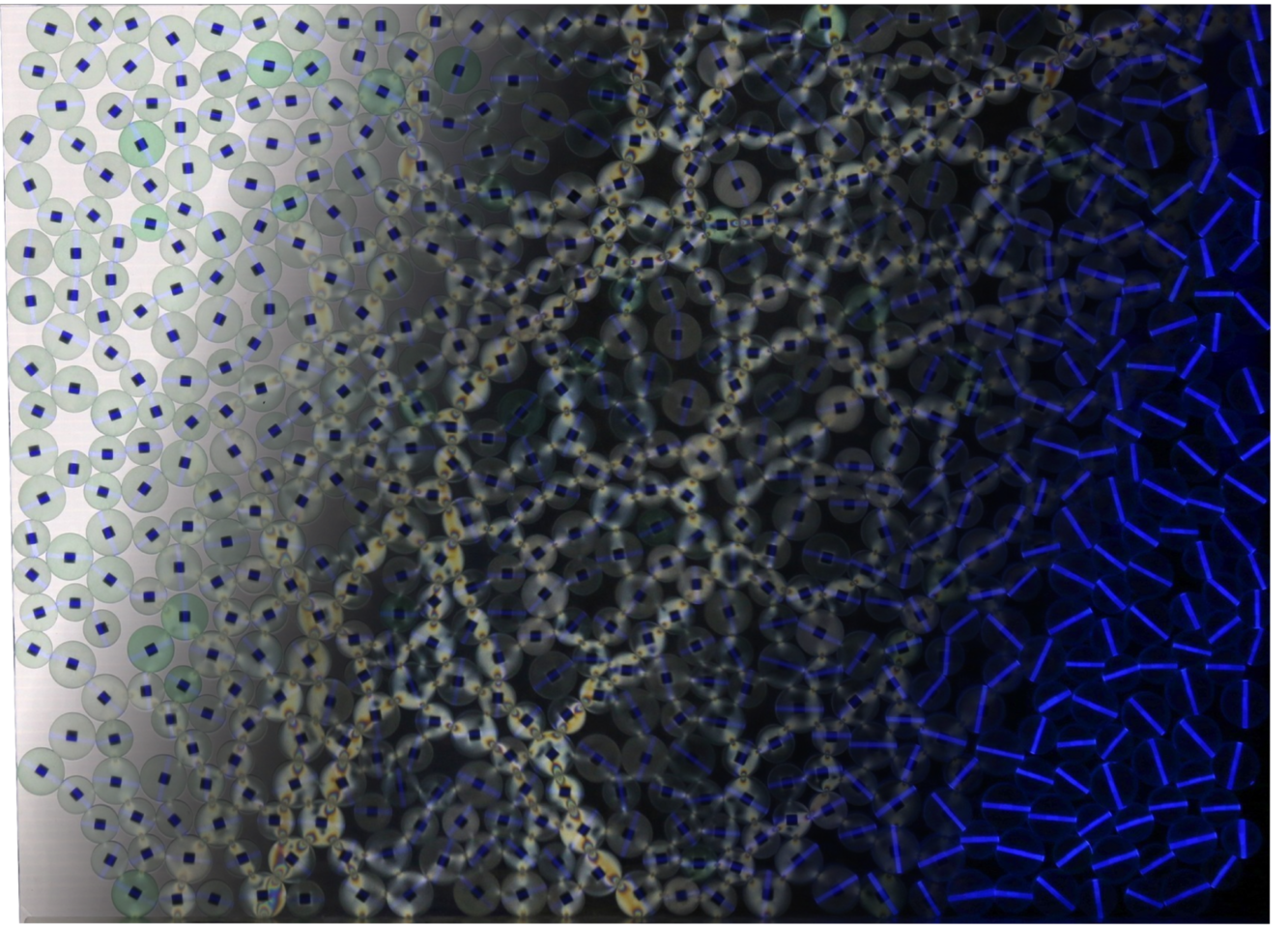}}
\caption{Composite view of a granular photoelastic system. On the left side, the grains are imaged with white backlighting. In the middle, the particles are viewed between crossed polarizers, revealing the force chains through photoelasticity. On the right side, the system is imaged from above with a UV light, revealing inked bars used to track particle rotation. Within each particle, an embedded cubic magnet is visible as a dark square \cite{cox2016_epl}.}
\label{fig_example}
\end{figure}

Most of the transparent objects we encounter are photoelastic: their degree of birefringence depends on the local stress  at each point in the material \cite{frocht1969_book,cloud1995_book}. This property can be used to visualize, and even quantitatively measure, what is usually invisible to our naked eye: the stress field. When such materials are subjected to an external load, and placed between crossed polarizing filters, each different region of the material rotates the light polarization according to the amount of local stress \cite{daniels2017_rsi}. This creates a visual pattern of alternating colored fringes (see Fig.~\ref{fig_example}) within the material which, on top of their aesthetic and pedagogical aspects, permits us to quantify the stress field within the material.

Photoelastimetry has its roots in engineering practice, where it was widely used to design parts before the rise of computational finite element methods \cite{cloud1995_book}. It also provided the first glimpse of the internal forces within granular materials, at first qualitatively \cite{wakabayashi1950_jpsj,dantu1957_proc,drescher1972_jmps,liu1995_sci} and later quantitatively \cite{howell1999_prl,majmudar2005_nat}. Today, it remains the most well-developed method for quantifying stresses \cite{amon2017_aip}, and methods for particle making \cite{cox2016_epl,bares2017_epj} and image post-processing \cite{daniels2017_rsi,kollmer2018_git,lantsoght2018_git} are under active development.

After two decades of quantitative efforts, the scientific successes of the photoelastimetry method are numerous. In Robert Behringer's group alone, it was responsible for identifying the erratic stress fluctuations in sheared granular matter \cite{howell1999_prl,bares2017_pre,zadeh2018_arx}, Green's function response \cite{geng2001_prl}, particle-scale anisotropy of the contact force networks \cite{majmudar2005_nat,zhang2010_gm}, shear jamming \cite{majmudar2007_prl,Bi2011,ren2013_prl,zheng2014_prl,wang2018_prl}, the dynamics of granular matter under impact \cite{clark2012_prl,lim2017_prl,zheng2018_pre0}, the Reynolds pressure, the Reynolds coefficient \cite{ren2013_prl}, and more. 

Far beyond the bounds of his laboratory at Duke University, the method has been used to examine particle shape dependence \cite{Zuriguel2008b}, identify interparticle contacts \cite{Lherminier2014}, observe sound propagation \cite{Shukla1991,Owens2011,Huillard2011}, test the validity of statistical ensembles \cite{puckett2013_prl,bililign_protocol_2019}, examine sensitivity to initial conditions \cite{kollmer2018_sm}, identify dilatancy softening \cite{Coulais2014}, measure force chain order parameters \cite{Iikawa2016}, and observe the effects of fluid flow \cite{Mahabadi2017}. In interdisciplinary efforts, photoelastimetry permits scientists to evaluate the grain-scale stresses caused by growing plant roots \cite{wendell2012_em,Kolb2012,bares2017_epj}, and examine situations relevant to faulting and earthquakes \cite{Daniels2008,Hayman2011,geller_stick-slip_2015,lherminier_continuously_2019}

The structure of the paper is as follows. First, we briefly review the physics of photoelasticity in \S\ref{sec_theory}; this section can be skipped for those only interested in qualitative uses of the method. In \S\ref{sec_particles}, we present various ways of fabricating photoelastic particles by  cutting, casting or printing, followed by imaging-techniques in \S\ref{sec_imaging}; these two section can stand alone for the creation of a demonstration apparatus. Finally, we present quantitative methods  in \S\ref{sec_postproc}. In all cases, additional information and technical specifications are provided on a wiki to which many of the paper authors have contributed \cite{all2019_wiki}.

\section{Photoelasticimetry theory} \label{sec_theory}

Photoelasticity arises from the birefringent properties of most transparent materials, in which the speed of light (via the index of refraction) depends on the polarization of the incident light wave. In some cases, such as glass and polymeric materials,  birefringence arises only when the material is subject to anisotropic stress, with the refractive indexes depending on the eigenvalues of the local stress tensor. As such, photoelasticity can provide measurements of the internal stress in the material.

\begin{figure*}
\centering 
\resizebox{0.7\hsize}{!}{\includegraphics{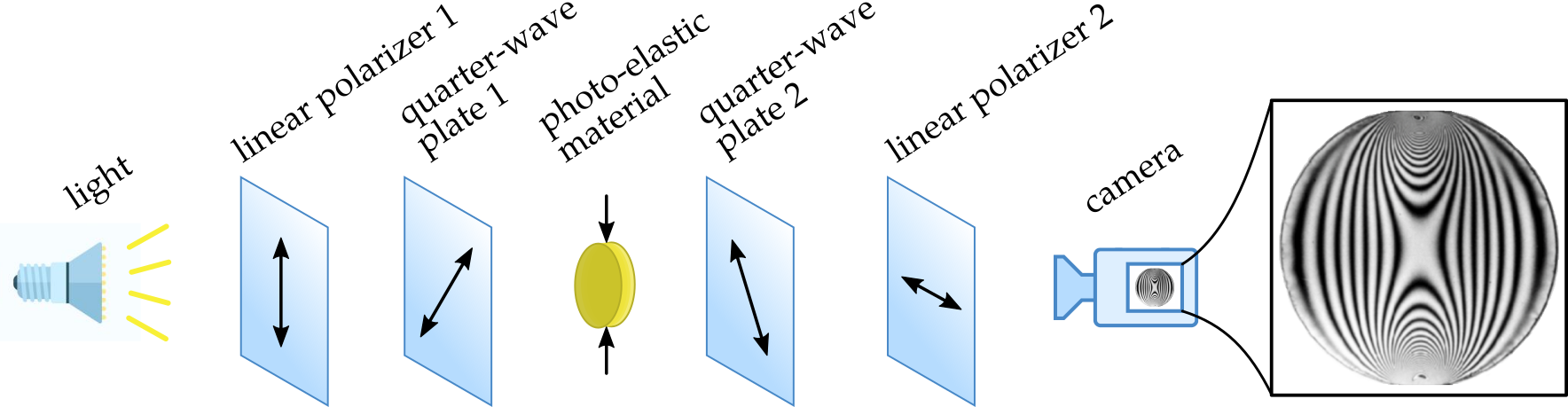}}
\caption{Schematic image of the photoelastic technique for a darkfield transmission polariscope: the combination of linear polarizer $1$ and quarter-wave plate $1$ (with a $\pi/4$ difference of principle directions), converts unpolarized light into circularly polarized light. This light passes through a uniaxially loaded photoelastic particle. A second combination of plate $2$ and linear polarizer $2$ (producing the opposite chirality) blocks any un-rotated light. At locations where there are anisotropic stresses in the photoelastic material, there will be different phase shifts for different components of the light wave; this phase shift changes the polarization of the light wave, causing these locations to appear bright in a camera image (or by eye). A sample image obtained with monochromatic light is given on the right hand side of the figure; each fringe is produced by an integer number of rotations of the polarization of light through a phase of $\pi$, as given by Eq.~\ref{eq_pol}.}
\label{fig_transmit}
\end{figure*}

Measurements are best taken using circularly (rather than linearly) polarized light, in order to provide isotropic measurements. Circularly polarized light is created using a linear polarizer followed by a quarter-wave ($\pi/2$) phase shift between the two orthogonal components, as shown in Fig.~\ref{fig_transmit}.
On the other side of the birefringent material, a circular polarizer with opposite polarity (the ``analyzer'') blocks any light that doesn't match its polarization. If the material is unstressed, there is no transmitted light and a dark image results. However, anyplace in the material where there is anisotropic stress, the wave components which are polarized along the two principle axes of the local stress tensor will travel with different speeds. This speed difference results in a relative phase shift for these two components of the wave, converting circularly polarized light to elliptically polarized light. As a consequence, a portion of the wave is not completely blocked by analyzing polarizer, and is therefore recorded as a bright region of the image. 

This property allows photoelasticity to be used to quantitatively measure local stress, via an inverse method. We begin by assuming that the relation between the local stress and the refractive index is linear. We consider the difference in transmission between the two principle axes: 
\begin{equation} 
    n_1-n_2=C(\sigma_1-\sigma_2),
    \label{eq_refract}
\end{equation} 
where $\sigma_1, \sigma_2$ are the two eigenvalues of the local stress tensor, and $n_1, n_2$ are the two refractive indices  in the corresponding directions. The material constant $C$ is known as the stress-optical coefficient. The relative phase shift of wave components in the eigendirections of the local stress tensor is determined from 
\begin{equation}
\alpha=\frac{2\pi C d}{\lambda}(\sigma_1-\sigma_2),
\end{equation}
with $\lambda$ the wavelength of the incident light and $d$ the distance traveled inside the material (its thickness). 
For this phase shift, the intensity of the wave that emerges from the analyzer is given by
\begin{equation} 
    I=I_0 \sin^2 \frac{\alpha}{2} =\sin^2 \left[ \frac{\pi C d}{\lambda}(\sigma_1-\sigma_2) \right].
    \label{eq_pol}
\end{equation} 
More details about these relationships and how to calibrate the material parameters for quantitative measurements can be found in \cite{Majmudar-thesis, daniels2017_rsi} and \S\ref{sec_postproc}.

Note that while Eq.~\ref{eq_pol} relates internal stress and image intensity for a single wavelength of light, the effect is also preset for a superposition of wavelengths (e.g. white light), as shown in Fig.~\ref{fig_example}. The difficulty is that inverting Eq.~\ref{eq_pol}, to infer local stress from light intensity,  requires the use of a single wavelength in order to be tractable. Even so, the  non-uniqueness of the solution due to the $\sin^2$ term makes the inversion problem challenging. \S\ref{sec_postproc} of this paper provides techniques for performing this task.

\section{Fabricating photoelastic particles} \label{sec_particles}

While many transparent materials have  photoelastic properties, only some of them have a large enough value of the stress-optical coefficient $C$ (see Eq.~\ref{eq_refract}) to produce a measurable effect under reasonable loads. Furthermore, it is necessary to pay special attention during the fabrication of a material in order not to create an object containing significant residual stresses. Therefore, the creation of a photoelastic granular material requires considering all of the following: (1)  the load you will to apply; (2) whether the shape of the particles matters; (3) how precisely you expect to make quantitative measurements, if at all; (4) what imaging method you will use; and (5) budget available. In this section, we explain several ways to meet these goals, ranging from simply cutting particles from pre-existing sheets up to casting bespoke particles. 

\begin{figure}
\centering \resizebox{0.8\hsize}{!}{\includegraphics{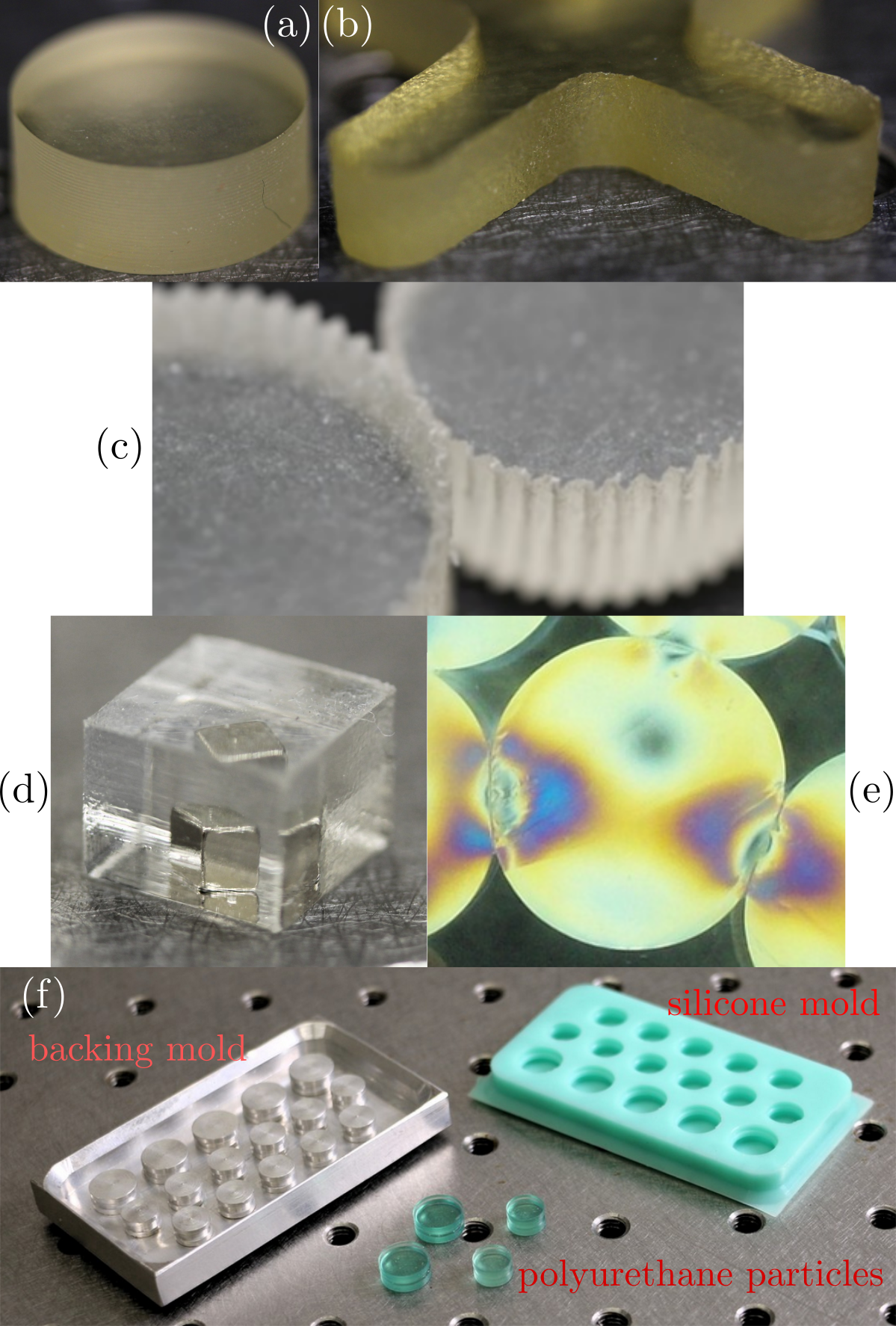}}
\caption{(a) Disc cut out of a photoelastic sheet \cite{precision_urethane} using a rotating cookie cutter.
(b) Cross-shaped particle cut out of a photoelastic sheet \cite{precision_urethane} using a computer-controlled milling machine.
(c) Geared particles cut out of a photoelastic sheet \cite{precision_urethane} using a computer-controlled waterjet.
(d) Cast polyurethane \cite{clear_flex} particle containing a magnetic inclusion. 
(e) Molded gelatin discs observed through crossed polarizers.
(f) Backing mold, mold and urethane cast photoelastic particles. All the particles are approximately 1 cm in size.}
\label{fig_particle}
\end{figure}

Before choosing the material for the particles, it is important to consider that the photoelastic signal is a periodic function of the stress (see Eq.~\ref{eq_pol}). The larger the deformation, the higher the stress, and the larger the number of fringes will be observed (see example in Fig.~\ref{fig_transmit}). Therefore, increased material stiffness is required for experiments with larger load, so that the fringes do not become denser than the resolution of the imaging system. Conversely, if the material is too stiff (or the loading too weak, or the thickness too small), then insufficient photoelastic signal will be observed. It is therefore advisable to perform some preliminary trials before committing to a large batch of particles. 

\subsection{Cutting sheets} \label{sec_cutting}

Historically, it was simplest to make photoelastic particles by simply cutting shapes out of a pre-existing sheet of photoelastic material. This could be a flat sheet of Plexiglas$^{\footnotesize{\texttrademark}}$ or a rubber, which is then cut with a bandsaw (the original Behringer particles), milling machine, spinning cookie cutter, or waterjet. The choice of material needs to match the experimenter's dual requirements of deformability and sensitivity ($C$ in Eq.~\ref{eq_refract}), to be compatible with the intended applied load. Most commonly, it is convenient to simply purchase sheets of transparent Vishay PhotoStress$^{\footnotesize{\texttrademark}}$ \cite{vishay} or polyurethane \cite{precision_urethane}, but for high loads Plexiglas$^{\footnotesize{\texttrademark}}$ or polycarbonate are also suitable choices. The stiffer materials can be machined using any appropriate tool, with residual stresses annealed out by heating them to just below their glass transition temperature. In all cases, a diversity of thickness, stiffness, and color are available. Sheets of thickness $1/4$'' and hardness $60$A are quite broadly-applicable, such that some particles of this type remain in use after two decades of use \cite{howell1999_prl,bares2017_pre}. 

If circular-shaped particles (discs) are desired, a custom-built ``cookie cutter'' tool can make numerous, identical particles from a single sheet with little waste \cite{daniels2017_rsi}. Both the rotation and downward cutting speeds have to be properly chosen as a function of the material not to induce residual stresses when cutting. An example of particle obtained with this method in given in Fig.~\ref{fig_particle}a, with slight horizontal marks made by the cutter. For more complicated shapes (see  Fig.~\ref{fig_particle}b), a computer-controlled  milling machine outfitted with a narrow-diameter mill can trace arbitrary outlines, again with care taken to minimize residual stresses. In this case, there is also some roughness at the outer surface. Finally, it is possible to create arbitrary shapes with waterjet cutting, first used by \citet{wendell2012_em} and further developed by \citet{wang2018_phd}. Depending on the skill of the operator, the resulting particles can have straight and smooth edges with few residual stresses, but can also sometimes leave a narrow channel at the start/end point of the cut. The narrow cutting width of the waterjet ($\sim 0.1$~mm) has the benefit of permitting very complex edges, as shown in Fig.~\ref{fig_particle}c.

\subsection{Casting particles} \label{sec_casting}

A second method to make photoelastic particles is to mold them directly, from such materials as polyurethane, gelatin, or nearly any other castable polymer or water gel. This allows an even larger diversity of complex, 3D shapes \cite{bares2017_epj} so long as a mold can itself be fabricated. Using this method, it becomes possible to tune the stiffness (via the controlled addition of crosslinking molecules) or to add inclusions \cite{cox2016_epl} as shown in Fig.~\ref{fig_particle}d. 

The first step of the casting method requires creating a backing mold: a positive relief of the desired particle shape. The backing mold can be machined or 3D printed of nearly any material stiff enough to maintain a shape. As shown in Fig.~\ref{fig_particle}f, this backing mold is then used to cast the final (reusable) silicone mold that makes the actual particles. A commercial silicone mold formulation such as MoldStar$^{\footnotesize{\texttrademark}}$ provides easy-to-use formulations \cite{mold_star}; we have found that 15 Slow fits most needs. 

Urethane is one popular choice of particle material \cite{bares2017_epj}. The commercial product ClearFlex$^{\footnotesize{\texttrademark}}$ \cite{clear_flex} is available in different stiffnesses ($50$A fits most needs), and it can be custom-tuned by varying the crosslinker ratio. For the benefit of particle-tracking (see \S\ref{sec_tracking}), it can be helpful to dye the clear urethane (see Fig.~\ref{fig_filter}). The product SoStrong$^{\footnotesize{\texttrademark}}$ \cite{so_strong} provides suitable dyes. Casting urethane requires some care to avoid the production of bubbles, compensate for material shrinkage, and develop fast enough work-flows; a number of helpful tricks from the community are shared at the online wiki \cite{all2019_wiki}.

Another popular choice is to use biological gels which are cheap and easy to use, but non-permanent. As shown in Fig.~\ref{fig_particle}e, gelatin has excellent photoelastic properties \cite{kilcast1984_jfs,lim2017_prl,workamp2017_epj}, as does agar or konjac \cite{tomlinson2015_oe} (with the later being less transparent). The stiffness of these materials is easy to tune by varying the ratio of gelling agent, and it is possible to achieve arbitrarily low elastic moduli. However, because biological gels are composed of water and a food source, they are vulnerable to drying, swelling,  and bacteria; so they are not stable over long times. Some groups have had success by crosslinking the gel \cite{damink1995_jms,workamp2016_rsi} when making particles \cite{workamp2016_rsi}. In this case, glutaraldehyde is directly added to the liquid gelatin preparation before molding or it can be diffused into the gelatin particles once they have gelled \cite{workamp2016_rsi} to increase the particle stiffness and stabilize the material.

\subsection{3D printing particles} \label{sec_3dprint}

Finally, when the particle shape is sufficiently complex that a backing mold cannot be made, it is possible to simply 3D print individual photoelastic particles. This choice is not as  straightforward as it might seem, since most of the available printing materials are non-transparent, porous, and contain residual stresses. However, optical fabrication processes (stereolytography) now allow for the production of semi-transparent photoelastic particles. Two available options are Durus White$^{\footnotesize{\texttrademark}}$ used by \citet{lherminier2016_rcf}, and VeroClear$^{\footnotesize{\texttrademark}}$  \cite{verro_clear} used by \cite{wang2017_sr}. The VeroClear has better clarity and optical response, but both are quite stiff and are therefore only appropriate for experiments with large loading stresses.

\section{Imaging methods} \label{sec_imaging}

In designing a photoelastic experiment, it is also necessary to consider the imaging method. The most appropriate choice will depend on considering the magnitude of the photoelastic response (\S\ref{sec_theory}) for the applied load, as well as desired degree of precision and the speed with which the images need to be collected to capture the dynamics. In this section, we elucidate several different methods that allow for adapted to  a variety of constraints. 

\subsection{Transmission vs. reflection imaging} \label{sec_polariscopes}

In \S\ref{sec_imaging} and Fig.~\ref{fig_transmit}, we considered a geometry in which the light is transmitted directly through the particles from a polarized light source, and then through a second filter (analyzer) before reaching the camera. In cases where the experiment will be optically-accessible from both sides of the particles, this is the easiest polariscope to construct, and has seen the most usage over the past several decades \cite{liu1995_sci,majmudar2005_nat,daniels2017_rsi}. Care must be taken when constructing the optical pathway of filters, that the polarizers be perfectly crossed and aligned exactly as shown in Fig.~\ref{fig_transmit}. This can be done either by purchasing two pairs of linear polarizers and quarter-wave plates \cite{polarizer} and doing the alignment by hand, or by purchasing a pair of left and right circular polarizers which are pre-aligned sandwhiches containing both components. In that case, care must be taken to orient the quarter-wave plate side of the polarizer towards the granular material so that light passes through the filters in the correct sequence. If the circular polarizers are pre-mounted inside a camera filter, their default configuration will be backwards from what is required for the polariscope shown in Fig.~\ref{fig_transmit}.

\begin{figure}
\centering \resizebox{0.75\hsize}{!}{\includegraphics{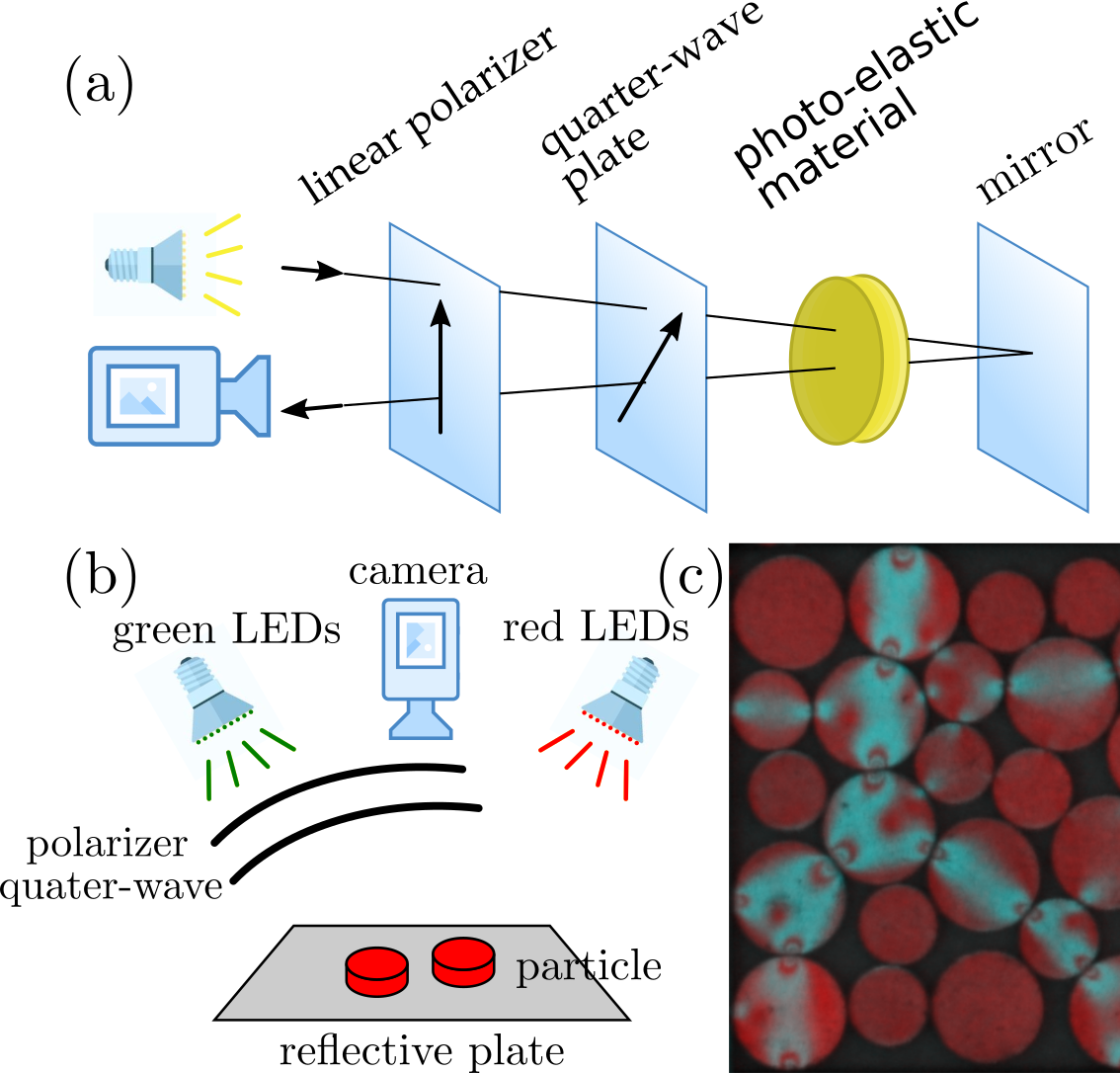}}
\caption{Schematic of the reflective photoelastic technique \cite{daniels2017_rsi}. (a) As for the transmission method (Fig.~\ref{fig_transmit}), the combination of  linear polarizer and quarter-wave plate converts unpolarized light into circularly polarized light. In this case, however, the reflector on the back side of the particle both creates the round-trip, and also reverses the polarization of the light. Therefore, there is only one polarizer that also serves as the analyzer.  (b) Sample reflective photoelastic set-up, illuminated with both a green polarized light and red unpolarized light. (c) Sample image from \cite{kollmer2018_sm}, recorded with a reflective polariscope of the type shown in (b). To render this red-green image accessible to more readers, the green channel has been copied into the blue channel.}
\label{fig_reflect}
\end{figure}

In some cases, the granular material may not be optically-accessible from both sides, for instance due to the loading mechanism or because the particles are resting on an opaque surface \cite{puckett2013_prl,zhao2017_epj}. It can therefore be desirable to create an optical setup in which the the apparatus is lit and imaged from the same side: this is a reflective polariscope. As in transmission polariscope, both the polarizer and the analyzer are circular, but now a single polarizer serves in both roles (see Fig.~\ref{fig_reflect}). Two successful options for reflecting the polarized light back through the granular sample are to rest the particles on a mirrored surface, or to coat all particles with a mirror-effect paint.  Details about how to construct such an apparatus are provided in \cite{daniels2017_rsi}.  

\subsection{Multi-wavelength imaging} \label{sec_wavelengths}

As described by Eq.~\ref{eq_pol}, quantitative stress measurements require monochromatic light measurements. In order to minimize the overlapping of photoelastic fringes at high stresses (as can seen in Fig.~\ref{fig_transmit}), it is important to work with monochromatic light. Furthermore, polarizers and quarter-wave plates are optimized for a given wavelength, usually green light. Therefore, a quantitative apparatus should be designed so that green light is used for photoelastic measurements, and other wavelengths are used for monitoring quantities such as particle positions and orientations.

One method, suitable for quasi-static dynamics, is to perform sequential imaging after each loading step \cite{ren2013_prl,bares2017_pre,zhao2017_epj}. For example, the camera takes an image (1) between crossed polarizers, to observe the photoelastic response; (2) without at least one of the polarizers, to detect particle positions; and (3) additionally illuminated by UV light, to monitor an additional characteristic such as particle orientation. An example of this type of imaging is shown in Fig.~\ref{fig_example} and Fig.~\ref{fig_disktrack}d, in which a bar drawn in UV-sensitive ink is visible.

\begin{figure}
\centering \resizebox{0.95\hsize}{!}{\includegraphics{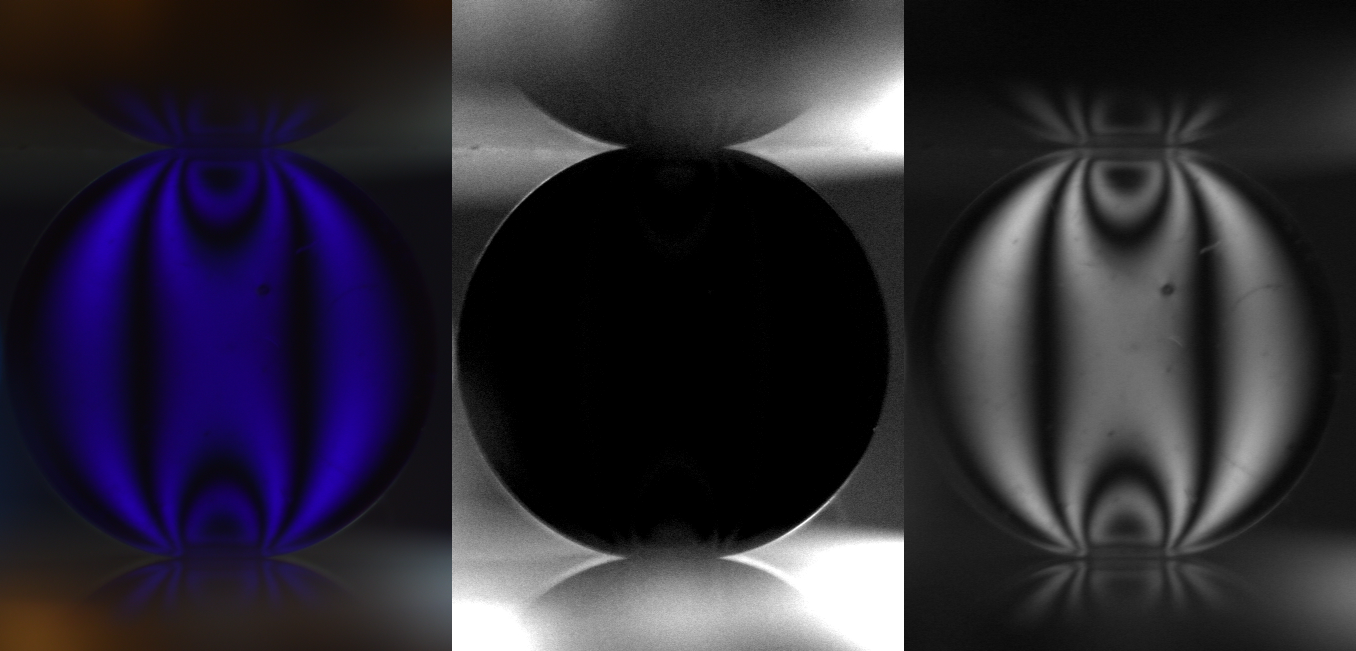}}
    \caption{A photoelastic particle dyed dark blue acts as wavelength filter. (Left) A color image of the particle illuminated by white light in a transmission polariscope can be split such that (middle) the green channel provides a brightfield image of a black disk, suitable for location detection, and (right) the blue channel provides a darkfield image of the photoelastic response.}
    \label{fig_filter}
\end{figure}

Another option is to measure these same fields simultaneously, by choosing to illuminate different properties using different wavelengths of light. Conveniently, color cameras already perform color-separation into red, green, and blue (RGB) channels. For example, the particle positions can be illuminated in unpolarized red light, while the photoelastic measurements are recorded by polarized green light, and a subset of tagged particles identified by UV-ink tags that emit blue light \cite{puckett2013_prl,kollmer2018_sm}. Such a setup is shown schematically in Fig.~\ref{fig_reflect}b, with a sample image in Fig.~\ref{fig_reflect}c. 

A third approach is to use dyed particles, which can then be illuminated with white light. In this case, the particles effectively act as a bandpass filter, and the passed wavelength can be used to make photoelastic measurements while blocked wavelengths provide position data. A particle dyed dark blue is shown to demonstrate this effect in Fig.~\ref{fig_filter}. These techniques allow a single image to provide both position detection and stress measurements at the same instant in time, making it possible to monitor the dynamics of a system.

\section{Analyzing images} \label{sec_postproc}

We have seen numerous examples of photoelastic images taken from a variety of geometries and lighting conditions; next, we examine several key methods of extracting quantitative information from such images. The level of precision available to the user -- whether particle position, orientation, or force --  depends on the focus and resolution of the images, the photoelastic sensitivity of the particles (parameter $C$ in Eq.~\ref{eq_refract}), and the magnitude of the applied load. In what follows, we will focus on the use of cylindrical (disk) particles.

\subsection{Tracking particle positions and orientations \label{sec_tracking}}

\begin{figure}
\centering \resizebox{0.95\hsize}{!}{\includegraphics{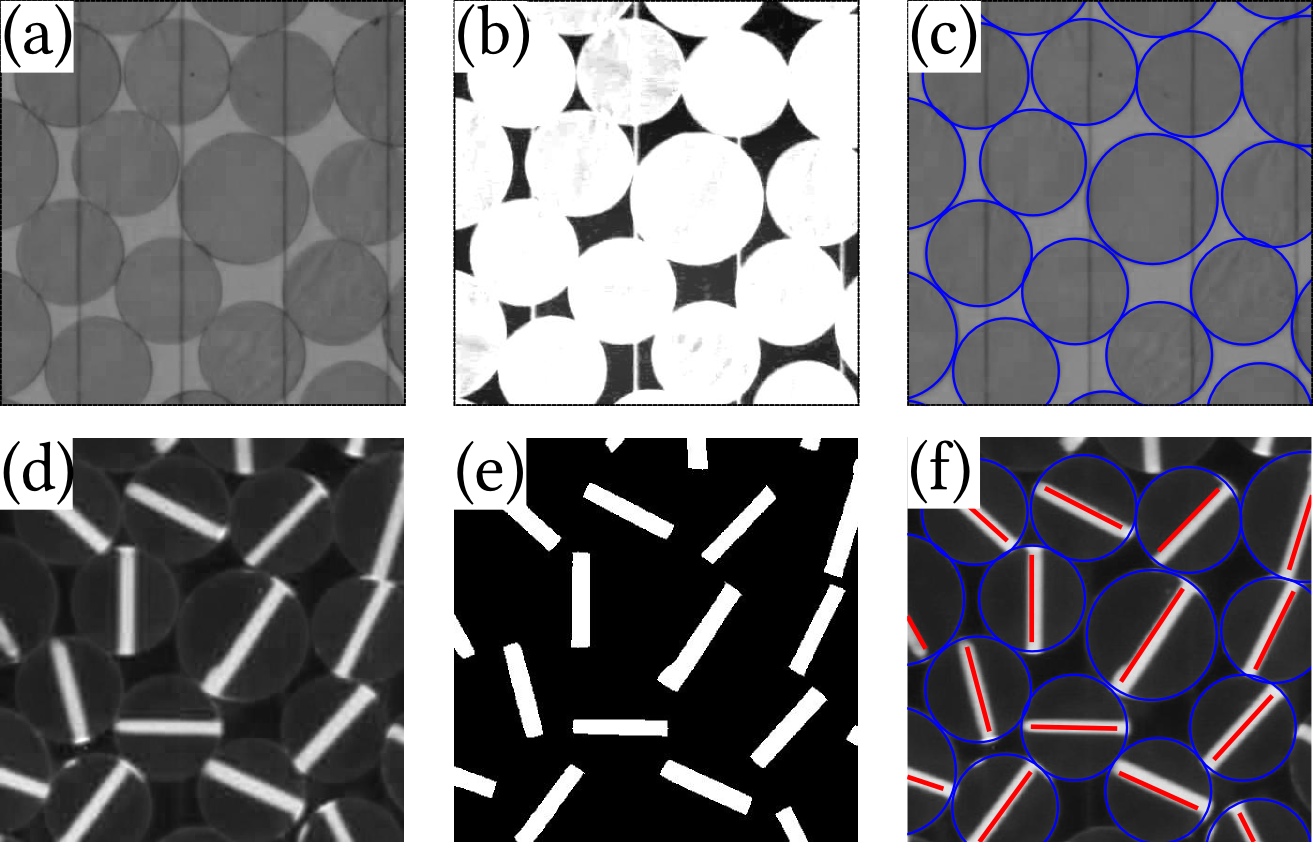}}
\caption{(a-c) Examples of locating particle-centers and measuring their diameters: (a) one channel of a originally-white light image, (b) enhanced contrast image with background noise removed, and (c) tracked particles shown as blue circles, found using the circular Hough transform method. (d-f) Examples of tracking particle orientations: (d) blue channel of a original UV light image, (e) enhanced contrast image with background noise removed, and (f) tracked orientations marked by red lines.}
\label{fig_disktrack}
\end{figure}

For particle detection, the first step is to produce a high-contrast (even binary) image to distinguish the pixels occupied by particles from those that are not. This is best done on whichever color-channel (red, green, or blue) has the  highest contrast between the particles and their background. It can be helpful to increase the contrast by filtering single high/low outlier pixel values (see Fig.~\ref{fig_disktrack}a), and also to apply a low-pass Fourier filter (see  Fig.~\ref{fig_disktrack}). These steps work for particles of any shape. 

Starting from a filtered image improves particle-finding methods, of which we will consider the two most convenient: convolutions \cite{Shattuck2015} and Hough transforms.  If the particles are not circular, then the convolution method is preferred over the Hough transform; it can also achieve higher precision but is more computationally-intensive. Image convolutions are performed between a pre-set image (``kernel'') of a single ideal particle (approximately $3/4$ the diameter of the smallest particle works well), and the original or binarized image. After performing the convolutions, binarizing the resulting image using a threshold of $99.5\%$ of the peak convolution value will result in a field of well-isolated peaks corresponding to the particle centers, with the area of each peak indicating that particle's diameter. For circular (non-binarized) images, it is possible to use a  circular Hough transform to locate the particle centers and sizes \cite{peng07_jcise, puckett2013_prl}. This transform works best on an image that has been prepared, via edge-detection, to isolate the circumference of particle. The algorithm uses a computationally-efficient voting process to identify the centers of such circles of a specified radius. A sample result of particle-detection via the Hough transform in shown in  Fig.~\ref{fig_disktrack}c.

From the the list of particle centers present in each image of a series, it is possible to track Lagrangian trajectories of those particles through space and time. Two efficient, accurate, and open-source algorithms are those of \citet{crocker1996_jcis} and \citet{BlairDufresne}. The main idea for both is that, given the positions for all detected particles at a given step, the aim is to select the closest possible new positions from the list of all detected particles in the next step. To this end, the algorithm considers all possible pairings of (old-new) positions, and selects the pairing that would result in the minimal total squared displacement. If a particle is missed over a small interval, the two halves of the trajectory later be re-attached. In cases where the image resolution was insufficient to reliably detect particle positions, it is still possible to use particle image velocimetry (PIV) to measure Eulerian flow fields for a series of images \cite{willert1991_ef}. 

Finally, it is possible to tag the particles with a contrasting stripe, in order to measure particle-rotation during dynamics. One common method, shown in Fig.~\ref{fig_disktrack}d-f, is to mark the top of each particle with a UV ink bar. This ink is not visible under white light, and therefore doesn't interfere with ordinary particle-tracking as described in the first part of this section. However, once illuminated by UV light alone, the ink bars become visible with high enough contrast to be recorded by a camera. The contrast is sufficient to binarize the image, and the position of each particle center can be used to conduct a least squares of the bar pixels in order to determine its orientation (see Fig.~\ref{fig_disktrack}f).

\subsection{Scalar measurements of force \label{sec_scalar}} 

\begin{figure*}
\centering \resizebox{0.7\hsize}{!}{\includegraphics{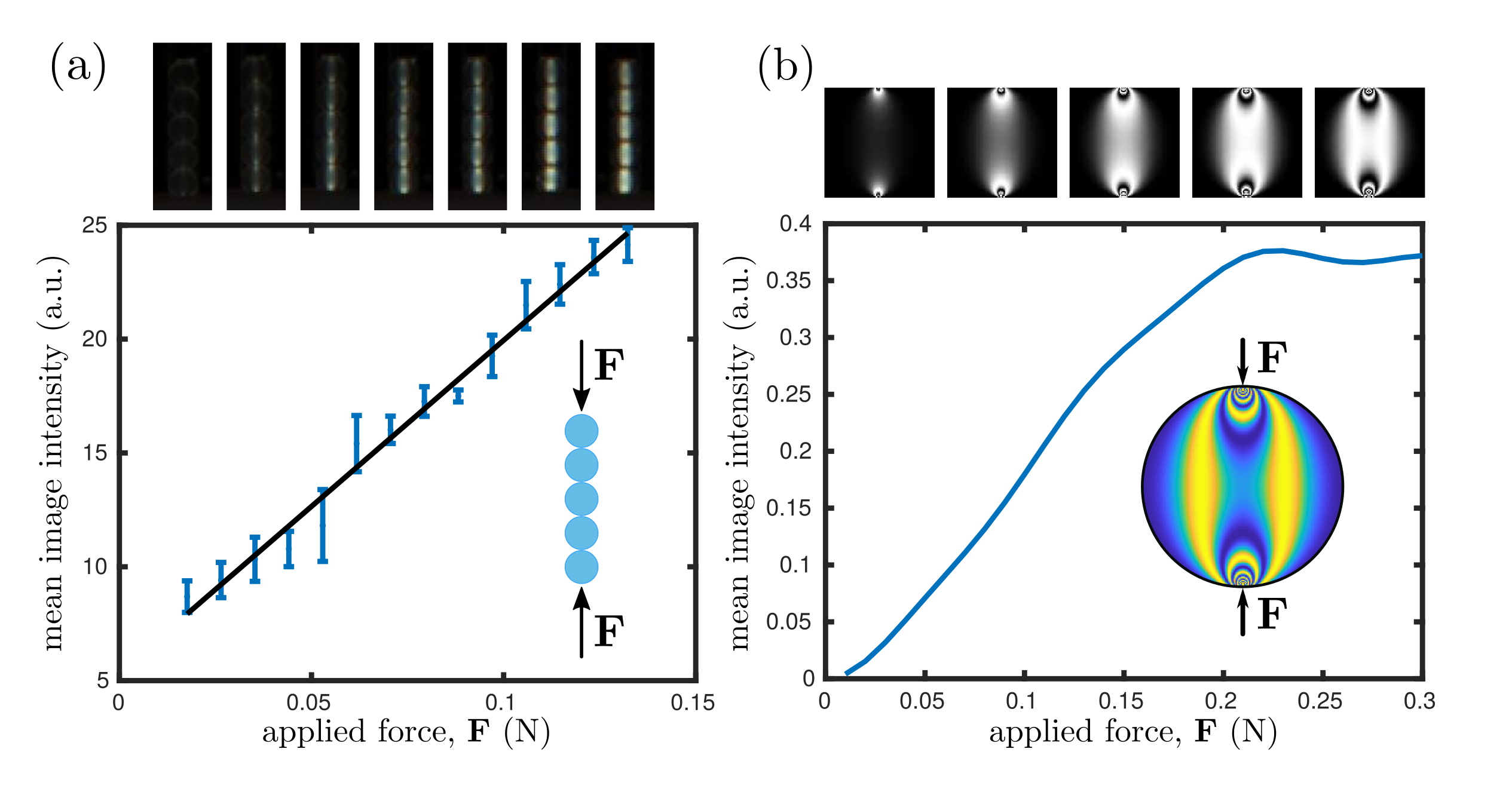}}
\caption{Calibration curve for mean image intensity on  (a) experiments and (b) numerical simulations of diametrically-loaded disks, as a function of applied force. Numerical simulultions are based on the stress field inside a single disk \cite{landau1986theory}, and the photoelastic response (Eq.~\ref{eq_pol}) is computed over a domain of $20 \times 20$ pixels, with $Cd=70$~m$^2$/N. Images corresponding to the different forces are shown on top of the diagram. The error bars are standard deviation on mean.}
\label{fig_fast}
\end{figure*}

For many experiments, it is desirable to make quick, low-resolution measurements of the forces on each particle, rather than solving the full inverse problem to obtain vector contact forces as will be described in \S\ref{sec_vector}. Scalar measurements can be the best choice for a variety of reasons: poor image resolution, dim lighting, fast dynamics, or a very large number of images. It is the easiest analysis when beginning a new project, as a tool for preliminary investigations, and remains popular because if its computational efficiency. 

The simplest form of scalar analysis is to use the mean light intensity within particles. For small forces, the mean image intensity increases linearly with the applied force, as shown in Fig.~\ref{fig_fast}a for a sample experiment under a known imposed load. By analyzing a small system such as a linear chain, it is possible to perform a calibration which provides a quantitative measure of the force on particle. As shown in Fig.~\ref{fig_fast}b for a numerical version of this calibration, there is a threshold force above which the mean light intensity plateaus and this method is no longer quantitative. This threshold corresponds approximately to the development of the first additional set of fringes. For applied forces within the linear regime, this method is valid. 

\begin{figure}
\centering \resizebox{0.95\hsize}{!}{\includegraphics{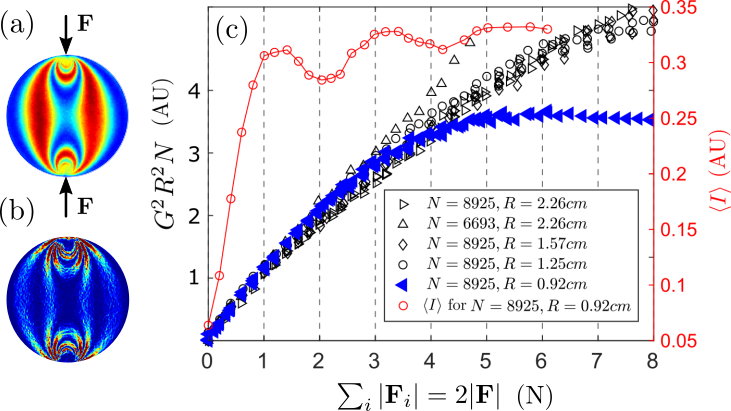}}
\caption{(a) Light intensity $I(i,j)$ of a diametrically loaded disk with force $\vec{F}$ between crossed polarizers. (b) The calculated $|\nabla I|^2$ distribution from (a). (a) and (b) are colored by the magnitude of $I$ and $|\nabla I|^2$ respectively. (c) Dependence of $G^2$ and averaged light intensity $\braket{I}$ on the contact forces for diametrically loaded disks with different $R$ (disk radius) and $N$ (number of pixels per meter) while keeping other experimental parameters the same. $\braket{}$ means averaging over pixels inside the disk. The black and blue data shows collapse of $G^2$ after proper rescaling using $R$  and $N$. The saturation forces for intensity and $G^2$ for an example run are about $1$~N and $4$~N respectively, showing that $G^2$ measures much larger range of forces. Data from \cite{zhao2019_npj}.}
\label{fig_g2}
\end{figure}

Once this threshold is crossed, the variation in intensity due to the fringes becomes a significant factor, allowing for an additional quantitative technique which accounts for the presence of the fringes. For images at high enough resolution to observe fringes, the force on the particle can be measured using the gradient of the image intensity field $I(i,j)$, known commonly as the $G^2$ technique.  It is defined as follows:
\begin{equation}
    \begin{array}{l}
        G^2 \equiv \braket{|\nabla I_{i,j}|^2} = \frac{1}{4}\langle [(I_{i-1,j}-I_{i+1,j})^2/4 \\ +(I_{i-1,j-1}-I_{i+1,j+1})^2/8+(I_{i,j-1}-I_{i,j+1})^2/4 \\
        +(I_{i+1,j-1}-I_{i-1,j+1})^2/8] \rangle
    \end{array}
\end{equation} 
where $i$ and $j$ refers to the row and column number of pixels and $\braket{\cdot}$ means averaging over pixels in the region of interest \cite{majmudar2006_phd}. Figs.~\ref{fig_g2} shows the distribution of light intensity $I$ (a) and its squared gradient $|\nabla I|^2$ (b) for a photoelastic disk under normal compression. 

This method has been used to measure both the average pressure for the whole image \cite{howell1999_prl,geng2001_prl,farhadi2014_prl,zheng2014_prl,farhadi2015_prl,cox2016_epl,daniels2017_rsi,zheng2018_pre0}, and at the particle scale \cite{clark2012_prl,Coulais2014,Iikawa2015_jpsj,bares2017_pre,wang2018_prl}. For bulk measurements, we observe that $G^2$ is a monotonic function of pressure, with some nonlinearity in the relationship  \cite{howell1999_prl}. For particle-scale measurements on disks, if applied forces are small enough such that photoelastic fringes can be clearly resolved, $G^2$ is proportional to the sum the individual vector contact forces $\sum_i|\vec{F}_i|$, where ${\vec F}_i$ are the 
contact forces on the disk \cite{zhao2019_npj} (see Fig.~\ref{fig_g2}a). Thus, where tangential forces are not large, $G^2$ approximately measures the particle scale pressure.

Importantly, $G^2$ will differ based on the material choice, the lighting condition and the image resolution. At particle scale, for an individual disk, the coefficient of proportionality $k=G^2/\sum_i|\vec{F_i}|$ is shown to be proportional to $I_0^2Cd/\lambda R^2N$ \cite{zhao2019_npj}, with $\lambda$ the light wavelength, $C$ the stress-optical coefficient, $d$ the disk height, $R$ its radius, $N$ the camera resolution defined as number of pixels per meter and $I_0$ the background light intensity. As an example, Fig.~\ref{fig_g2}c plots the collapse of the linear part of $G^2$ rescaled using $N$ and $R$ as functions of $\sum_i|\bf{F_i}|$ for diametrically loaded disks (see Fig.~\ref{fig_g2}a) with other parameters kept the same. The dependence of $G^2$ on $R$ is important for analyzing poly-disperse systems. Fig.~\ref{fig_g2}c also compares the dependence of $G^2$ and intensity $I$ on the contact forces for a given test, showing that the range of force that $G^2$ can measure is $4$ to $5$ times larger than that of $I$. Note that the linear dependence of $G^2$ value for individual disk on contact forces will differ for other particle shapes \cite{zhao2019_npj}.

\subsection{Vector measurements of force \label{sec_vector}}

\begin{figure}
    \centering \resizebox{0.95\hsize}{!}{\includegraphics{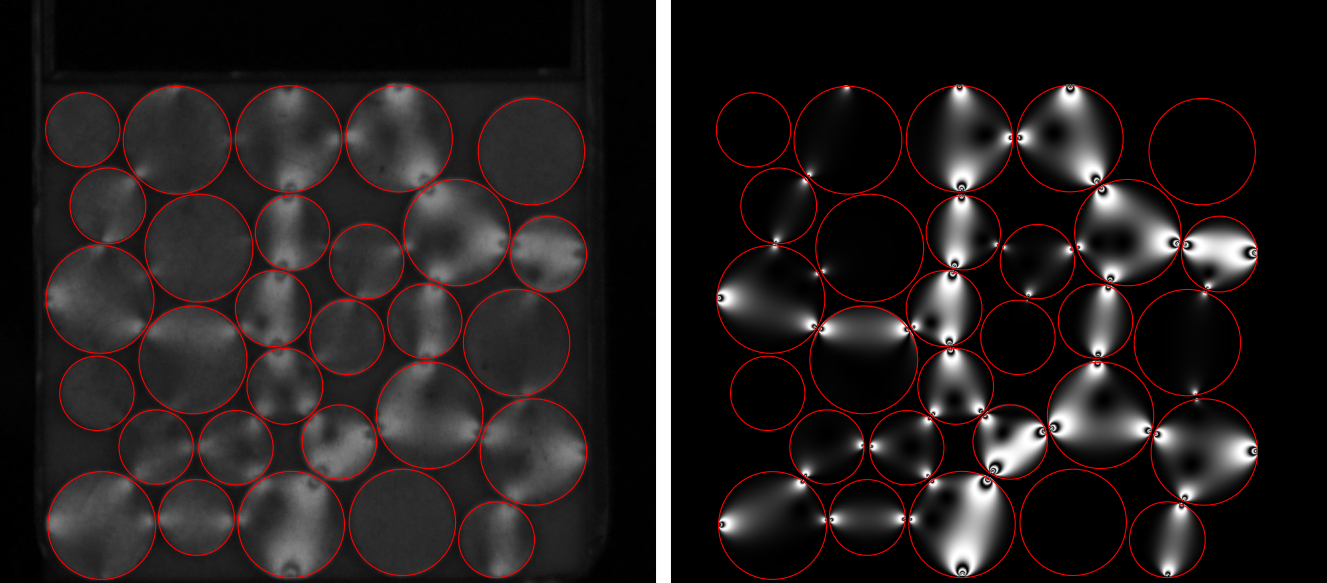}}
    \caption{Experimental photoelastic response (left side)  and numerically generated counterpart (right side) after the optimization of all ${\vec F}_i$ contact forces. Data from \cite{kollmer2018_git}.}
    \label{fig_FM}
\end{figure}

The detailed pattern of light and dark fringes in $I(i,j)$ is set by the local stress values through Eq.~\ref{eq_pol}. Therefore, it is possible to use a known stress field (computed from the set of vector contact forces ${\vec F}_i$) to predict a light intensity field. Examples of such calculations are shown in Fig.~\ref{fig_fast}b. In order to determine the values for each ${\vec F}_i$ {\it from} the fringe pattern requires performing the inverse of this process, as illustrated in Fig.~\ref{fig_FM}. Because Eq.~\ref{eq_pol} cannot be directly inverted due to the $\sin^2$ non-monotonic function, the numerical estimation of the set of ${\vec F}_i$ must be done via an optimization process. This is achieved in a sequential process: 
\begin{enumerate}
    \item Detect the particle positions (see \S\ref{sec_tracking})
    \item Estimate the total and individual contact force magnitudes on each particle (see \S\ref{sec_scalar})
    \item Optimize ${\vec F}_i$ on each particle according to a set of constraints
\end{enumerate}

Note that the second step contains two parts. It is usually possible to use the $G^2$ method to estimate the total force, but this may not be possible for the individual contacts, depending on the image resolution. If the quality of the photoelastic image is good enough, each contact force can be similarly estimated from calculating $G^2$ in the vicinity of each contact \cite{daniels2017_rsi,kollmer2018_git}. This method allows for the removal of contacts that do not transmit loads, and for the inverse problem to be performed on individual particles. However, the contact-scale step is not always possible for lower-resolution images, for example in the case of stirred granular media \cite{Lantsoght2016_phd}. Instead, the optimization step (below) must also include a process for testing multiple different contact-configurations in which the total force magnitude is taken to be equally-distributed in the initiation of the optimizer \cite{lantsoght2018_git}. 

The third step proceeds to optimize the vector contact forces ${\vec F}_i$, starting from the estimated values in steps 1 and 2, in order to achieve a photoelastic response as close as possible to the image intensity $I(i,j)$. The input (initial guess) and output (optimized) parameters are therefore the set of all contact force magnitudes and orientations. For cylindrical, linearly-elastic particles subjected to multiple contact forces, the stress-field can be approximated by an analytical expression \cite{timoshenko1970_book,daniels2017_rsi}. 

Combining this stress field with Eq.~\ref{eq_pol} creates a reconstructed image of each particle which can be evaluated against the measured image. The chosen optimization algorithm works by evaluating the agreement between these two images -- the measured $I(i,j)$ and the numerically-generated $I_n(i,j)$ -- and modifying the values of ${\vec F}_i$ to bring them into closer agreement. The similarity between $I, I_n$ can be computed with such quantities as the mean squared error which is quick to calculate, or the structural similarity index \cite{wang2004_IEEE} which better-accounts for the the image structure. One commonly-used  optimization protocol is Levenberg-Marquardt optimization \cite{daniels2017_rsi}.

No matter the choice of similarity index or optimization protocol, it is beneficial to have the initial guesses for forces to be of highest quality possible in order to have the optimizer convergence to a valid result. When the initial guesses are too far from the correct values, the optimizer may land in a local minimum and not escape. Ideally, a single run of optimization is sufficient to determine ${\vec F}_i$ for all particles, but in practice it can be beneficial to run the process multiple times. When the optimizer is having difficulty converging, for instance due to poor-quality initial guesses for the contact forces, it is possible to propagate already-identified values of ${\vec F}_i$ to their reciprocal force on the adjacent particle. This improves the initial guesses for the next particle, and this process can be repeated multiple times, sequentially propagating information through the packing. At the end of the optimization process, all contact forces (magnitude and orientation) have been determined for each of the particles in the system. At this stage, it is still possible to iteratively make additional improvements, for instance by checking for consistency with Newton's Third Law, or the  equilibrium condition on each particle \cite{daniels2017_rsi}.

\section{Outlook} 

We close with a summary of the newest developments in photoelastimetry: taking these techniques and applying them to faster dynamics, non-circular particles, and three-dimensional systems.

\begin{figure}
\centering \resizebox{0.75\hsize}{!}{\includegraphics{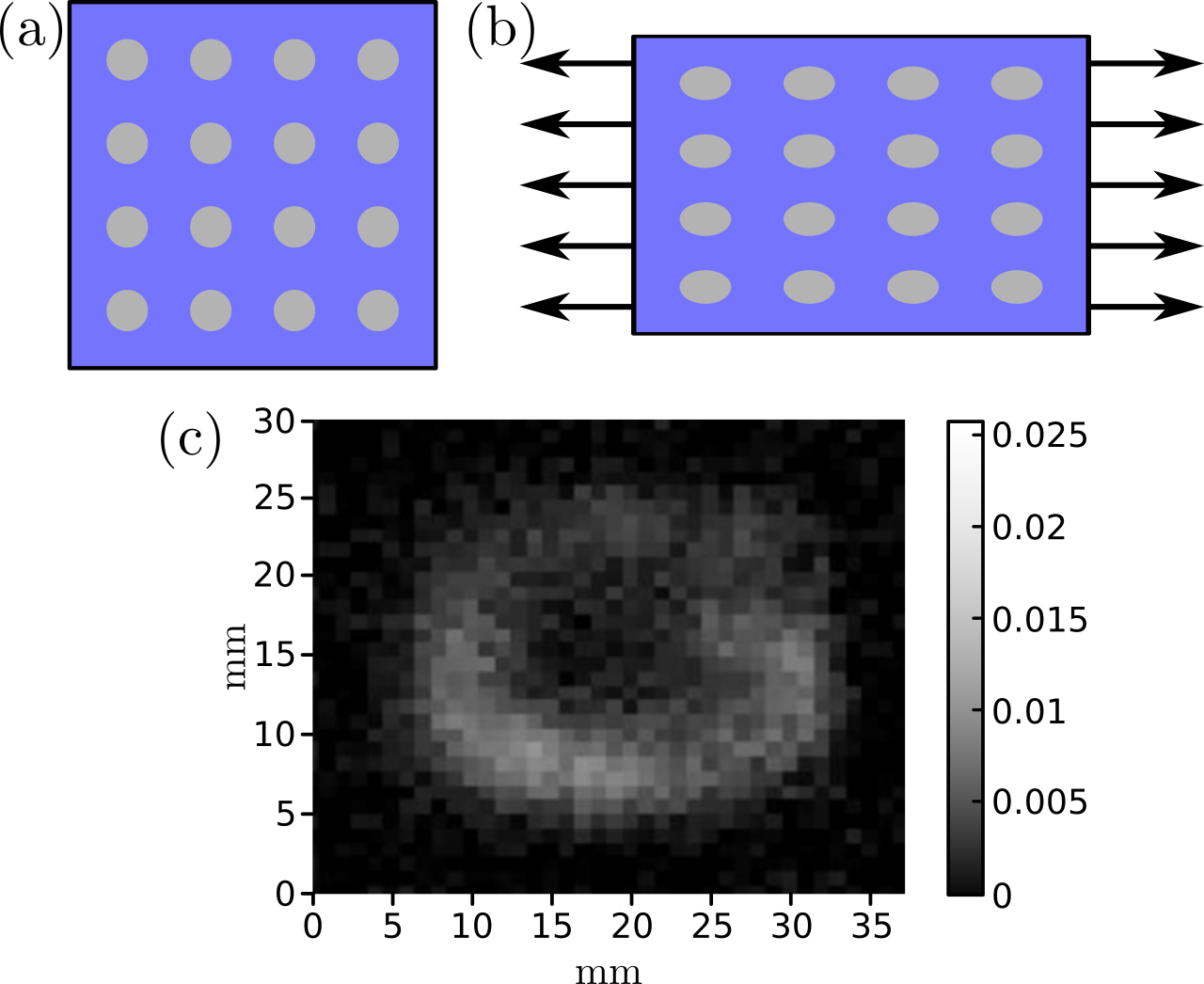}}
\caption{Schematic of a metamaterial exhibiting terahertz frequency strain birefringence using reversible, elastic meta-atoms in unstressed (a) and stretched (b) situations. (c) Spatial map of the photoelastic strain measured for a terahertz metamaterial. The stretched sample was placed between crossed polarizers and raster scanned through a narrow terahertz beam at the resonance frequency that was detected by a single broadband detector. The grayscale represents the unnormalized cross polarimetric signal indicating the regions experiencing the greatest strain.}
\label{fig_THz}
\end{figure}

Photoelastic measurements in fully three-dimensional systems is challenging, since a curved particle simultaneously acts as a lens. Nonetheless, it is possible to obtain semi-quantitative information about the stress state of the system using birefringent spheres \cite{liu1995_sci,yu_monitoring_2014}. A promising route for 3D studies is therefore to use  terahertz photoelasticity, where the wavelength of light is hundreds of microns. Although early measurements confirmed the translucence of photoelastic materials at terahertz frequencies \cite{mahon2006_oc,heimbeck2011_oe}, the first attempts at measuring strain birefringence failed because the microscopic displacements of atoms were too small compared to the wavelength of the radiation. However, it is possible to develop  metamaterials whose meta-``atoms'' exhibit greater displacements. An example of such a material is shown in Fig.~\ref{fig_THz}ab. The basic principle, recently demonstrated \cite{everitt2019_aom}, is that an applied stress reversibly distorts the shape of the meta-atoms, and this anisotropy is detected when the material is placed between crossed polarizers and illuminated from above by a terahertz generator. To make truly 3D measurements, this system can be combined with terahertz holographic imaging  \cite{mahon2006_oc,heimbeck2011_oe}. 

Another challenge in applying the techniques described in \S\ref{sec_postproc} to more realistic situations, is to allow their application to particles that are non-circular or cohesive, such as commonly arise in geophysical and industrial contexts. For circular particles with cohesion (i.e. electrostatic interactions, liquid bridges, chemical bonds), the inverse methods described in \S\ref{sec_vector} likely still apply. However, the key technical challenge for non-circular particles is that the inverse problem is only well-specified for the case of circular particles, where the trivial surface geometry (all tangential forces are torques, all normal are central) dramatically simplify the formulation of the inverse solution to Eq.~(\ref{eq_pol}). New mathematical solutions are required to created a more general formulation to apply to particles of arbitrary shape.

Another persistent challenge is to move from quasi-static experiments to fully dynamic. Previously, this was accomplished by performing semi-quantitative force analysis, as described by the $G^2$ method in \S\ref{sec_scalar}. Recently, advances in high-speed imaging have made it possible to record images with sufficient brightness and resolution to perform photoelastic analysis on continuously-avalanching flows, and thereby obtain vector contact forces via adaptations of the methods in \S\ref{sec_vector} \cite{thomas_photoelastic_2019}. This process is enabled by measuring the particle positions directly from a single movie of photoelastic images, in order to capture as much light as possible. Furthermore, because the particles are accelerating, it is not possible to use particle-scale force-balance as a constraint in determining the vector contact forces.

Finally, computational advances in inverting the photoelastic images would improve the applicability of photoelastimetry to all of the problems presented here. Better algorithms for avoiding unintended-minima and the ability to simultaneously solve for all vector contact forces in the system (a massive set of constraints) would improve the reliability of these methods.

It is our hope that the recent proliferation of publications of photoelastimetric techniques in open formats \cite{all2019_wiki,lantsoght2018_git,kollmer2018_git} will spur development in all of these new directions.

\section*{Acknowledgements}

We would like to thank R\'{e}my Mozul for his technical support with the wiki \cite{all2019_wiki}. Several conversations and collaborations have led to sharing the techniques described in this paper. We are grateful to Bernie Jelinek and Richard Nappi for sharing their technical knowledge about photoelastic material cutting. The outlook section contains insights gained from Chris M. Bingham, Willie J. Padilla, Anthony Llopis and Nan M. Jokerst (recent work on terahertz photoelasticity), and from Nathalie Vriend and Amalia Thomas (fast-imaging photoelasticity). 

Finally, we thank the late Robert Behringer for his kindness, his depth of knowledge gained from developing photoelastic techniques for two decades, and his stimulating attitude towards every new generation of scientists passing through his laboratory. This review article is a product of his excellent mentorship.

\bibliographystyle{unsrtnat}
\bibliography{biblio,Photoelasticity}

\end{document}